\documentclass[11pt,twoside]{article}
\usepackage{./asp2014}

\aspSuppressVolSlug
\resetcounters

\begin{document}

\title{Intermediate-Mass Black Holes in Globular Cluster Systems}

\author{J. M. Wrobel,$^1$ J. C. A. Miller-Jones,$^2$ K. E. Nyland,$^3$
  and T. J. Maccarone$^4$
\affil{$^1$National Radio Astronomy Observatory, P.O. Box O, Socorro,
  NM 87801, USA; \email{jwrobel@nrao.edu}}
\affil{$^2$International Centre for Radio Astronomy Research, Curtin
  University, GPO Box U1987, Perth, WA 6845, Australia;
  \email{james.miller-jones@curtin.edu.au}}
\affil{$^3$National Radio Astronomy Observatory, Charlottesville, VA
  22903, USA; \email{knyland@nrao.edu}}
\affil{$^4$Department of Physics and Astronomy, Texas Tech University,
  Box 41051, Lubbock, TX 79409-1051, USA; \email{
thomas.maccarone@ttu.edu}}}

\paperauthor{J. M. Wrobel}{jwrobel@nrao.edu}{0000-0001-9720-7398}{NRAO}{}{Socorro}{NM}{87801}{USA}
\paperauthor{J. C. A. Miller-Jones}{james.miller-jones@curtin.edu.au}{0000-0003-3124-2814}{Curtin University}{}{Perth}{WA}{6845}{Australia}
\paperauthor{K. E. Nyland}{knyland@nrao.edu}{0000-0003-1991-370X}{NRAO}{}{Charlottesville}{VA}{22903}{USA}
\paperauthor{T. J. Maccarone}{thomas.maccarone@ttu.edu}{0000-0003-0976-4755}{Texas
  Tech University}{}{Lubbock}{TX}{79409}{USA}

\begin{abstract}
  Using the Next Generation Very Large Array (ngVLA), we will make a
  comprehensive inventory of intermediate-mass black holes (IMBHs) in
  hundreds of globular cluster systems out to a distance of 25~Mpc.
  IMBHs have masses $M_{IMBH} \sim 100-100,000~M_\odot$.  Finding them
  in globular clusters would validate a formation channel for seed
  black holes in the early universe and inform event predictions for
  gravitational wave facilities.  Reaching a large number of globular
  clusters is key, as Fragione et al.\ (2018) predict that only a few
  percent will have retained their gravitational-wave fostering IMBHs.
\end{abstract}

\section{Scientific Motivation}

Theory suggests that globular clusters (GCs) of stars can host
intermediate-mass black holes (IMBHs) with masses $M_{IMBH} \sim
100-100,000~M_\odot$ \citep{mil02,gur04,por04,gie15,mez17}.  Finding
IMBHs in GCs would validate a formation channel for seed black holes
(BHs) in the early universe \citep[e.g.,][]{vol10,sak17}; populate the
mass gap between the well-understood stellar-mass BHs and the
well-studied supermassive BHs \citep[e.g.,][]{kor13,tet16a}; and test
if scaling relations between stellar systems and their central BHs
extend into poorly-explored mass regimes \citep[e.g.,][]{gra15}.
Also, the GC system of a typical galaxy contains several hundred GCs
\citep{har16}.  Thus studying such a system could constrain the mass
spectrum of IMBHs in its GCs and, related, the ability of its GCs to
retain their IMBHs and foster gravitational wave (GW) events
\citep[e.g.,][]{hol08,fra18}.  These properties could vary from galaxy
to galaxy, so many GC systems should be studied.  Importantly, a GC
system with a low fraction of IMBHs at present could be linked to a
high rate of GW events in the past.

To search for IMBHs in GCs, one looks for evidence that the IMBHs are
affecting the properties of their GC hosts.  For GCs in the Local
Group, a common approach is to use optical or infrared data to look
for the dynamical signatures of IMBHs on the orbits of stars in the
GCs.  Such sphere-of-influence studies have a contentious history
\citep[e.g.,][and references therein]{bau17,mez17}, even leading to
differing IMBH masses when using the orbits of stars or of radio
pulsars in the same GC \citep[e.g.,][]{gie17,per17}.  Having
3-dimensional velocity fields should improve the dynamical searches
\citep{dru03}.  Still, they are fundamentally limited by shot noise
due the small number of orbits influenced \citep[e.g.,][]{van13}.  The
dynamical searches are also susceptible to measuring high
concentrations of stellar remnants rather than an IMBH
\citep[e.g.,][]{den14}.

An independent approach that bypasses these issues is to look for the
accretion signatures of IMBHs in GCs \citep[][and references
  therein]{mac16}.  This approach leverages on decades of studies of
the signatures of accretion onto both stellar-mass and supermassive
BHs \citep[][and references therein]{fen16}.  Here, we apply a
synchrotron radio model to search for the signatures of low rates of
accretion onto IMBHs in entire GC systems.  The model was developed
for GCs in the Local Group \citep{mac04,mac08,mac10,str12} and is
summarized in Section~2.  We have used the US National Science
Foundation's Karl G.\ Jansky Very Large Array \citep[VLA;][]{per11} to
demonstrate the feasibility of a radio search for IMBHs in one nearby
GC system \citep{wro16} and summarize that effort in Section~3.  In
Section~4 we demonstrate the breakthrough role that the Next
Generation VLA \citep[ngVLA;][]{sel17} will have in searching for
IMBHs in hundreds of GC systems hosted in nearby galaxies.  We close,
in Section~5, by linking these searches to related studies using
facilities contemporary with the ngVLA.

\section{Synchrotron Radio Model}

We invoke a semi-empirical model to predict the mass of an IMBH that,
if accreting slowly from the tenuous gas supplied by evolving stars,
is consistent with the synchrotron radio luminosity of a GC
\citep[][and references therein]{str12}.  Following \citet{str12}, we
assume gas-capture at 3\% of the Bondi rate \citep{pel05,mer07} for
gas at a density of 0.2 particles~cm$^{-3}$ as measured by
\citet{fre01}, and at a constant temperature of 10,000~K as justified
by \citet{sco75}.  We also assume that accretion proceeds at less than
2\% of the Eddington rate, thus involving an advection-dominated
accretion flow with a predictable, persistent X-ray luminosity.  (An
IMBH accreting at higher than 2\% of the Eddington rate would enter an
X-ray-luminous state \citep{mac03} and be easily detectable in
existing surveys.  But no such X-ray sources exist in Galactic GCs.)
We then use the empirical fundamental-plane of BH activity
\citep{mer03,fal04,plo12} to predict the synchrotron radio luminosity.
The radio emission is expected to be persistent, flat-spectrum,
jet-like but spatially unresolved, and located near the dynamical
center of the GC.

\begin{figure}[!htb]
\centering\includegraphics[scale=0.5]{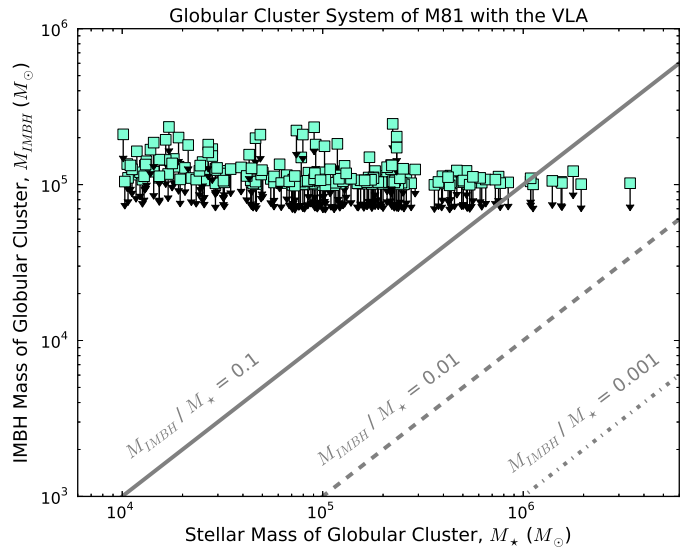}
\caption{3$\sigma$ upper limits to the mass of the IMBH, $M_{IMBH}$,
  according to the Strader et al.\ (2012) model, as a function of the
  stellar mass, $M_\star$, for probable GCs in M81 (Wrobel et al.\
  2016).  The grey diagonal lines of constant $M_{IMBH} / M_\star$
  convey fiducial ratios of IMBH mass to stellar mass.}\label{f1}
\end{figure}

From parameter uncertainties, \citet{str12} estimate that the IMBH
mass associated with a given radio luminosity could be in error by
0.39 dex, thus a factor of 2.5.  To improve the robustness of such
masses, one could fold in more recent data, especially regarding the
highly uncertain gas-capture parameter.  It could also be profitable
to examine the model's underlying framework.  For example, guided by
\citet{sco75}, the model assumes that the classic Bondi flow toward a
point mass, the IMBH, is isothermal.  Yet the classic Bondi accretion
rate would be lower if the flow could be argued to be adiabatic
\citep{pel05,per17}.  On the other hand, models for an isothermal flow
toward an IMBH embedded in a realistic GC potential appear to achieve
higher accretion rates than classic Bondi flows that are isothermal
\citep{pep13}.

\section{Globular Cluster System of M81 with the VLA}

We searched for the radiative signatures of IMBH accretion from 206
probable GCs in M81 \citep{wro16}, a spiral galaxy at a distance of
3.63~Mpc \citep{fre94}.  Forty percent of the probable GCs are
spectroscopically confirmed, with the balance deemed to be good GC
candidates \citep{nan11}.  Our search used a four-pointing VLA mosaic
at a wavelength of 5.5~cm and a spatial resolution of 1.5~arcsec
(26.4~pc).  It achieved 3$\sigma$ upper limits of $3 \times
(4.3-51)~\mu$Jy for point sources in individual GCs, depending on
their location in the mosaic.  Weighted-mean image stacks yielded
3$\sigma$ upper limits of $3 \times 0.43~\mu$Jy for all GCs and $3
\times 0.74~\mu$Jy for the 49 GCs with stellar masses $M_\star \gtrsim
200,000 M_\odot$ based on the \citet{nan11} photometry.  We applied
the \citet{str12} synchrotron model to predict the IMBH mass,
$M_{IMBH}$, consistent with a given luminosity at 5.5~cm.  Figure~1
shows upper limits on $M_{IMBH}$ and $M_\star$ for the individual GCs.

For 7 GCs in M81 the ratios of their IMBH masses to stellar masses,
$M_{IMBH} / M_\star$, appear to be less than 0.03-0.09.  These upper
limits are competitive with the value of 0.02 for the
five-times-closer G1, the only extragalactic GC with an IMBH inferred
by spatially resolving its sphere of influence on cluster stars
\citep{geb05}. (Note, though, that G1's X-ray and radio properties
suggest a ratio of less than 0.01 \citep{mil12}.)  Also, the M81
stacks correspond to mean IMBH masses of less than 42,000~$M_\odot$
for all GCs and less than 51,000~$M_\odot$ for the 49 GCs with high
stellar masses.  The VLA is thus making inroads into the
difficult-to-observe regime of IMBHs in extragalactic GCs.  Further
significant progress demands deeper radio observations of individual
GCs in M81 and in other nearby GC systems.  Such efforts will
admittedly have poorer mass sensitivities than possible for individual
Galactic GCs \citep[e.g.,][]{str12}.  But an extragalactic observation
can capture many GCs and its mass sensitivity can be improved by
stacking those GCs.

\section{Globular Cluster Systems with the ngVLA}

We consider using Band 3 of the ngVLA \citep{sel17} to examine
globular cluster systems in the local universe.  Band 3 has a central
frequency of 17~GHz and a bandwidth of 8.4~GHz.  We approximate its
central wavelength as 2~cm.  The \citet{har13} compilation of GC
systems involves 422 galaxies.  The distribution of the galaxies'
distances shows two peaks.  A minor peak contains tens of galaxies
with distances out to 10~Mpc.  A major peak contains hundreds of
galaxies with distances between 10 and 25~Mpc.  A typical galaxy's GC
system holds several hundred GCs spread over an effective diameter of
a few tens of kpcs \citep{har16,kar16}.  Notably, the major peak in
distance contains tens of thousands of GCs in total.  Figure~2 shows a
potentially rich GC system at 23~Mpc \citep{bro14}.

\begin{figure}[!htb]
\centering\includegraphics[scale=1.7]{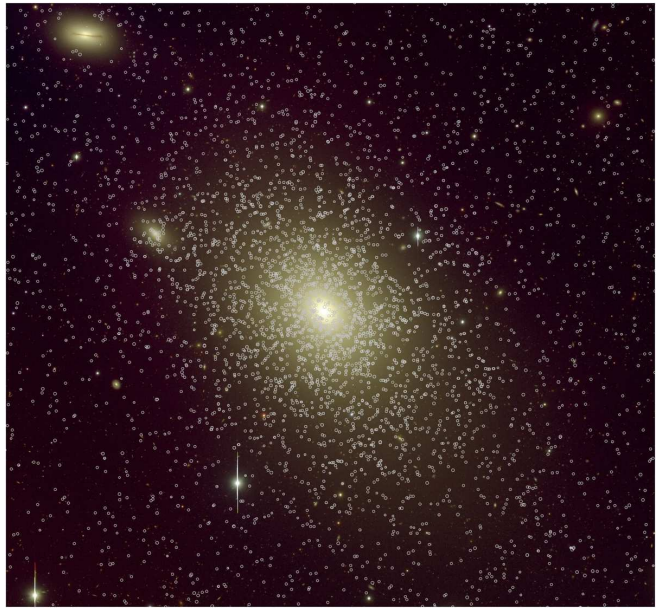}
\caption{GC system of the early-type galaxy NGC\,4365 at a distance of
  23~Mpc.  The small circles mark the GC candidates in the inner
  18~arcmin $\times$ 17~arcmin of a {\em gri} Suprime-Cam image.
  1~arcmin = 6.7~kpc.  From Brodie et al.\ (2014).}\label{f2}
\end{figure}

We applied the \citet{str12} synchrotron model to predict the
luminosity at 2~cm as a function of the mass, $M_{IMBH}$, of a
putative IMBH in a GC.  The associated point-source flux densities,
$S_{2cm}$, were then derived for GCs at distances of 10 and 25~Mpc.
In Figure~3, the sloping lines show how to convert from $S_{2cm}$ to
$M_{IMBH}$ for the two distances, while the vertical lines show
3$\sigma$ detections with the ngVLA, assuming tapered and robust
weighting, with integrations of 1, 10 and 100~hours \citep{sel17}.  At
higher signal-to-noise ratios, the wide frequency coverage could test
the flat-spectrum prediction, as well as raise flags about
steep-spectrum contaminants.

At 10~Mpc the synchrotron model predicts a flux density of $S_{2cm} =
0.35~\mu$Jy from an IMBH of 83,000~$M_\odot$.  The ngVLA can make a
3$\sigma$ detection with a 10-hour integration and a tapered,
robustly-weighted resolution of 100~mas (5~pc).  This resolution
matches the half-starlight diameter of a GC \citep{bro06}.  From
\citet{sel17}, the field of view (FOV) of the ngVLA is a circle of
diameter 3.4~arcmin (10~kpc) at full width half maximum.  Most of a GC
system can therefore be encompassed in a few FOVs.  Each FOV can
simultaneously capture many GCs.  Undetected GCs can also be stacked.
A stacking performance as for M81 (Section~3) can improve the IMBH
mass sensitivity by a factor of two.  At 25~Mpc the synchrotron model
predicts a flux density of $S_{2cm} = 0.35~\mu$Jy from an IMBH of
163,000~$M_\odot$.  The ngVLA can make a 3$\sigma$ detection with a
10-hour integration, localize the source to the GC, and encompass most
of the GC system in a few FOVs.  Each FOV can simultaneously capture
many GCs.  Stacking can also be done, and is expected to reach the
mass sensitivity mentioned above for an individual GC at 10~Mpc.

\begin{figure}[!htb]
\centering\includegraphics[scale=0.5]{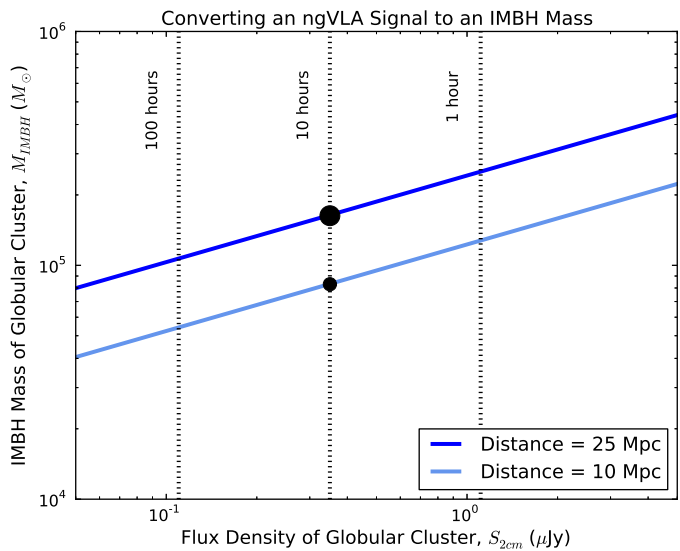}
\caption{ngVLA signals, $S_{2cm}$, from IMBH masses, $M_{IMBH}$, in
  GCs at distances of 10 and 25~Mpc, according to the Strader et
  al.\ (2012) model.  The small and big black dots highlight 3$\sigma$
  mass sensitivities at 10 and 25~Mpc, respectively, after 10 hours on
  target.  Hundreds of GC systems have distances between 10 and
  25~Mpc, and hold tens of thousands of GCs in total.  Reaching large
  numbers of GCs is key, as Fragione et al.\ (2018) predict that only
  a few percent will have retained their IMBHs.}\label{f3}
\end{figure}

Regarding possible radio contaminants in extragalactic GCs, guidance
comes from radio studies of X-ray binaries in the Galaxy and M31.
Dozens of persistent radio emitters are known in the Galaxy, and their
range of radio luminosities suggests negligible contamination beyond
the Local Group \citep[e.g.,][]{tet16b}.  One radio-flaring X-ray
binary has been identified in M31 \citep{mid13}.  A radio analog of it
could be detected out to 25~Mpc, but it would fade after a few months
so not be mistaken for a persistent emitter.  Three Galactic X-ray
binaries have similarly strong radio emission.  In two cases this is
likely related to massive donar stars, absent from GCs.  The third
case, GRS\,1915+105, is an unusually long-lived transient that is both
radio and X-ray bright \citep{fen04}.  An analog of GRS\,1915+105
would be detectable at both wavelengths out to 25~Mpc, but the
fundamental plane of BH activity would unmask it as a stellar-mass BH.
Also, when searching tens of thousands of GCs, we should beware of
possible radio contamination from background source populations.
Simulated source counts near 2~cm suggest that star forming galaxies
will dominate at $\mu$Jy levels \citep{wil08}.  But such galaxies have
steep specta and finite sizes, so will not be mistaken for IMBHs that
have flat spectra and are point-like.

{\em In summary, with its sensitivity, bandwidth, spatial resolution,
  and FOV, the ngVLA at 2~cm will efficiently probe IMBH masses in
  hundreds of GC systems out to a distance of 25~Mpc.}

\section{The ngVLA and Its Contemporary Facilities in the 2030s}

\noindent{\underbar{Gravitational Wave Facilities.}}  \citet{fra18}
explored the fate of primordial GCs, each born with a central
IMBH. They modelled the evolution of the GCs in their host galaxy, and
of the IMBHs undergoing successive, GW-producing mergers with
stellar-mass BHs in the GCs.  For primordial GCs that survived to the
present day, they found that a few percent retained their IMBHs and
the balance lost their IMBH when a GW recoil ejected it from the GC
host.  Once ejected, the IMBHs are no longer able to foster GW events.
They used these results to predict GW event rates for the {\em Laser
  Interferometer Space Antenna} \citep[{\em LISA};][]{ama17} and
Europe's Einstein Telescope \citep{hil11}.  The rates for the latter
facility also apply to the US's similarly-scoped Cosmic Explorer
\citep{abb16}.  IMBHs with masses between 1000 and 10,000~$M_\odot$
yielded mergers at rates that could be detected by all three GW
facilities.  IMBHs with masses $\gtrsim 10,000~M_\odot$ yielded
mergers at rates that could be detected only by {\em LISA}.  If the
ngVLA searches do not find the expected mix of IMBHs in GC systems, it
could challenge the framework underlying the GW predictions.  ngVLA
searches for point-like emission from IMBHs would be easy to conduct
during Early Science, notionally starting in 2028.  Guided by such early ngVLA
results, \citet{fra18} could revisit their GW predictions.

{\underbar{Electromagnetic Wave Facilities.}}  A key science driver
for extremely large telescopes (ELTs) in the 30-m class is to measure,
at a distance of 10~Mpc, a BH mass as low as 100,000~$M_\odot$ by
spatially resolving its sphere of influence in its GC host
\citep{do14}.  For example, if the Infrared Imaging Spectrometer
(IRIS) \footnote{\url{https://www.tmt.org/page/science-instruments}}
on the Thirty Meter Telescope (TMT) can achieve the diffraction limit
of 18~mas at 2~$\mu$m, then this approach will yield a sample of IMBHs
in GCs out to a distance of 10~Mpc.  A sphere-of-influence study with
IRIS must be done one GC at a time, a shortcoming that makes it
expensive to inventory the range of IMBH masses in a galaxy's GC
system.  The TMT's Infrared Multi-object Spectrometer (IRMS)
\footnote{\url{https://www.tmt.org/page/science-instruments}} in
spectroscopy mode will have a FOV of 2.0~arcmin $\times$ 0.6~arcmin.
This being a tenth of the ngVLA FOV, a sphere-of-influence study with
IRMS would require ten pointings to cover one ngVLA pointing.
Regardless of the situation at 10~Mpc, the ELT approach cannot reach
the hundreds of GC systems with distances between 10 and 25~Mpc.

The {\em Chandra} X-ray mission and its proposed successors, {\em
  Lynx}
\footnote{\url{http://cxc.harvard.edu/cdo/cxo2lynx2017/index.html}}
and the {\em Advanced X-ray Imaging Satellite}
\footnote{\url{http://axis.astro.umd.edu}}, feature spatial
resolutions of 300 to 500~mas. These will suffice to roughly localize
X-ray sources to GCs out to a distance of 25~Mpc.  But an X-ray--only
search for the accretion signatures of IMBHs in GCs will be hindered
by confusion from X-ray binaries in GCs \citep[e.g.,][]{jos17}.
Specifically, X-ray--only detections of GCs cannot discriminate
between X-ray binaries and IMBHs.  Fortunately, the empirical
fundamental-plane of BH activity \citep{mer03,fal04,plo12} implies
that the persistent radio emission from IMBHs is expected to be
several hundred times greater than that from X-ray binaries.  Thus
ngVLA imaging can be used to separate X-ray detections into bins for
X-ray binaries and for IMBHs.  X-ray binaries are known to be
time-variable in both the radio and X-ray bands, so this radio--X-ray
synergy would be strengthened by simultaneous observations with the
ngVLA and the X-ray mission.

The deployment baseline of SKA1-Mid \citep{dew15,bor17} will offer a
spatial resolution of 57~mas at 3.3~cm with uniform weighting.  This
resolution suffices to search for the accretion signatures of IMBHs in
GCs with declinations south of 10 degrees.  But as only 67 SKA1-Mid
antennas will be available at 3.3~cm, the effective collecting area of
that telescope will only be about that of the current VLA, which is
insufficient for our purposes.

Many of the GC systems in \citet{har13} will need further optical
imaging before finding charts are available for all of their GCs.
Such images can mostly be acquired with current ground-based
telescopes in the 4-m class \citep[e.g.,][]{har14} or 8-m class
\citep[e.g.,][]{bro14}.  But for some particularly confused regions,
it may be necessary to obtain images from space-based telescopes.  For
example, the Near Infrared Camera on the {\em James Webb Space
  Telescope} will offer spatial resolutions of 80 and 160~mas in its
short- and long-wavelength channels, respectively
\footnote{\url{https://jwst.stsci.edu/files/live/sites/jwst/files/home/instrumentation/technical\
    documents/jwst-pocket-guide.pdf}}.

\acknowledgements The National Radio Astronomy Observatory is a
facility of the National Science Foundation, operated under
cooperative agreement by Associated Universities, Inc.



\end{document}